\documentclass[a4paper,10pt]{article}
\usepackage{graphicx}
\usepackage{epsfig}
\usepackage{jcappub}
\usepackage[utf8]{inputenc}
\usepackage{multirow}
\usepackage{graphicx,amsmath,amsfonts,amssymb,slashed,upgreek,color,caption}
\begin{document}

\title{Clustering GCG: a viable option for unified dark matter-dark energy?}
\author{Sumit Kumar and  Anjan A Sen}
\affiliation{Centre For Theoretical Physics, Jamia Millia Islamia, New Delhi-110025, India}

\emailAdd{sumit@ctp-jamia.res.in, aasen@jmi.ac.in}

\abstract{We study the clustering Generalized Chaplygin Gas (GCG) as a possible candidate for dark matter-dark energy unification. The vanishing speed of sound ($c_{s}^2 = 0$) for the GCG fluid can be obtained by incorporating higher derivative operator in the original K-essence Lagrangian. The evolution of the density fluctuations in the GCG+Baryon fluid is studied in the linear regime. The observational constraints on the model are obtained using latest data from SNIa, $H(z)$, BAO and also for the $f\sigma_{8}$ measurements. The matter power spectra for the allowed parameter values are well behaved without any unphysical features.}
\date{today}

\maketitle

\section{Introduction}
After many years of theoretical and observational efforts, we know only $5\%$ of the energy budget of our Universe. Our knowledge for the rest $95\%$ is still incomplete. Through the gravitational interactions at astrophysical as well as cosmological scales we know that $(1/3)$rd of this unknown component is relatively massive, non-interacting as well as non-relativistic with negligible pressure. This is known as "dark matter" (DM) \cite{dm}. Existence of dark matter can explain the flat rotation curves for spiral galaxies  and it is also necessary to form the structures in our Universe. The remaining $(2/3)$rd of the unknown component is practically massless and also has the exotic negative pressure.  This is known as "dark energy" (DE) \cite{de} and is responsible for the late time acceleration of the Universe. Although there have been many theoretical attempts to explain DM and DE, we still do not have a robust model for DM and DE that can pass all the theoretical and observational tests.

Attempts have also been made to build models that can explain both DM and DE with a single field. These are popularly known as``unified models for DM and DE"  (UDME) in the literature. In these models, one introduces a single fluid that behaves as DM as well as DE at relevant time and length scales. One widely studied model for such a scenario is the ``Chaplygin Gas" (CG) \cite{ cg} and its subsequent generalisation which is known as the ``Generalised Chaplygin Gas" (GCG) \cite{gcg} (please see \cite{udm} for other model building attempts for UDME). Although these models were consistent with observations related to the background cosmology, it was subsequently shown that they produce unphysical features in the matter power spectrum in the form of huge oscillations or exponential blow-ups which are not seen in the observed matter power spectrum \cite{sand} (See also \cite{finelli} for other studies on GCG related to structure formation and CMB).  The main reason for such discrepancy is the behaviour of sound speed through CG/GCG fluid. In the early universe, when CG/GCG behaves as DM, the sound speed through the fluid vanishes and the CG/GCG clusters like non-relativistic dust. But in  late times when CG/GCG starts behaving like DE, the sound speed through the fluid becomes large resulting unphysical features in the matter power spectrum. To avoid such unphysical features, CG/GCG has to behave just like the concordance $\Lambda$CDM model. This was a blow not only to CG/GCG but practically to all attempts to unify the DM and DE (see \cite{soln} for attempts to overcome this problem).

There is a well established field theory model for GCG where one can write a K-essence Lagrangian for GCG \cite{gcg}. For a particular value of the GCG parameter $\alpha$ (see next section for details), this reduces to the well known Dirac-Born-Infeld (DBI) Lagrangian for d-brane in a (d+1,1) space time \cite{jackiew}. In recent past, Creminelli et al. \cite{crem} have shown that for the K-essence action, one can add a specific higher derivative term in the action that keeps the background equation of motion unaltered.  But for the perturbed universe, this extra higher derivative operator can result vanishing sound speed in the fluid. In such a scenario, the pressure perturbation vanishes and the K-essence will cluster at all scales like the non-relativistic matter. Hence these are called ``clustering quintessence" model. In recent past, several investigations have been done to study the observational signatures of the clustering quintessence models\cite{vern}, \cite{other}.

Given the fact that GCG as a UDME model failed because of the large sound speed through the fluid during the DE domination, it will be interesting to study the consequences of clustering GCG model. Because GCG fluid has an underlying K-essence type field theory, one can extend the results by Creminelli et al. \cite{crem} safely to GCG fluid and can keep its sound speed negligibly small thereby avoiding any unphysical features in the matter power spectrum.

In this work we study the clustering GCG  (CGCG) as a viable option for UDME candidate. We study the density perturbations in the GCG fluid ( which acts as an UDME). We also keep the baryons together with the GCG fluid. We confront the model with observational data from SNIa, BAO, H(z), $f\sigma_{8}$, and show that matter power spectra for the allowed parameter values are well behaved without any unphysical behaviour or unwarranted features.

\section{Generalised Chaplygin Gas}

We start with a Lagrangian \cite{gcg}:

\begin{equation}
{\cal L} = -A^{1/(1+\alpha)} \left[ 1 - X ^{(1+\alpha)/{2\alpha}}\right]^{\alpha/(1+\alpha)},
\end{equation}

\noindent 
where $A$ and $\alpha$ are constant and $X = g^{\mu\nu}\phi_{,\mu}\phi_{,\nu}$. With $\alpha = 1$ the Lagrangian reproduces the famous DBI Lagrangian. One can calculate the energy density and pressure for this field using the relations:

\begin{eqnarray}
p &=& {\cal L},\nonumber\\
\rho &=& 2 Xp_{,X}  -p,
\end{eqnarray}

\noindent
and it is straightforward to show that $\rho$ and $p$ are related by the following relation \cite{gcg}:

\begin{equation}
p = -\frac{A}{\rho^{\alpha}}.
\end{equation}

\noindent
The corresponding fluid that satisfies this equation of state is known as GCG \cite{gcg}. For $\alpha = -1$, GCG behaves like a normal fluid with a constant equation of state $w = -A$. $\alpha = 1$ corresponds to the original ``Chaplygin Gas" (CG) equation of state \cite{cg}. Using equation (2.3) in the energy conservation equation $\dot{\rho} + 3 H (\rho +p) = 0$, one can calculate the form of the energy density for GCG in a FRW background which is given by:

\begin{equation}
\rho_{g}= \left[A + \frac{B}{a^{(3(1+\alpha))}}\right]^{1/(1+\alpha)}.
\end{equation}

\noindent
Here $B$ is an arbitrary integration constant and we set $a_{0} =1$ at present. From now on subscript ``0" stands for values at present. One can write the above expression in a slightly different way as:

\begin{equation}
\rho_{g} = \rho_{g0}\left[A_s + (1 - A_s)a^{-3(1 + \alpha)}\right]^{1/(1 + \alpha)},
\end{equation}

\noindent
where we define $\rho_{g0}=(A+B)^{1/(1+\alpha)}$ and $A_{s} = A/(A+B)$.  We consider the ranges $0\leq A_{s} \leq 1$ and $-1 < \alpha \leq 1$ for our subsequent calculations. This ensures the energy density $\rho_{g}$ for the GCG  is well behaved for the entire evolution history that we consider. We can further write the expression for the equation of state GCG fluid as:

\begin{equation}
w_{g} = -\frac{A_{s}}{A_{s} + (1-A_{s})a^{-3(1 + \alpha)}}.
\end{equation}

It is straightforward to verify that at early times, $a<<1$, $\rho_{g} \sim a^{-3}$ and $w_{g} \sim 0$, hence GCG behaves like a non-relativistic matter, whereas for $a >>1$, $\rho_{g} = constant$ and $w_{g} \sim -1$ and GCG becomes a cosmological constant. In between, the equation of state $w_{g}$ makes a transition from $w_{g}=0$ to $w_{g} <0$ and starts behaving like a dark energy at late times. Epoch of the transition depends on the choice of the parameters $A_{s}$ and $\alpha$. This behaviour of $w_{g}$ makes GCG a suitable candidate for UDME at least for the background cosmology.  

We write down the expression for the Hubble parameter for GCG as an UDME in presence of baryons assuming a flat FRW universe:

\begin{equation}
\frac{H^2}{H_0^2} = \Omega_{b0}a^{-3} + (1-\Omega_{b0})(A_s + (1 - A_s)a^{-3(1 + \alpha)})^{1/(1 + \alpha)}.
\end{equation}

\noindent
Here $\Omega_{b0}$ is the present day density parameter for baryon. In this equation, the second term on right hand side is due to GCG which includes both cold dark matter (CDM) and DE. The $\alpha = 0$ limit corresponds to $\Lambda CDM$ model and  we can easily identify $(1-\Omega_{b0})A_s$ to be the $\Omega_{\Lambda 0}$ and $(1-\Omega_{b0})(1-A_s)$ to be  the $\Omega_{c 0}$ (where "c" stands for CDM) as in concordance $\Lambda CDM$ cosmology. 

To be a successful model for UDME, GCG should also mimic the inhomogeneous universe as in the $\Lambda$CDM model. For this it is necessary that GCG clusters similarly as CDM at all scales. This depends on the sound speed through the GCG fluid. The sound speed through the GCG fluid can be calculated using the corresponding expression for the K-essence field as obtained by Garriga and Mukhanov earlier \cite{gm}:

\begin{equation}
c_{s} ^2 = \frac{p_{,X}}{\rho_{,X}} = -\alpha w_{g}.
\end{equation}

In the early time, when GCG behaves like a non-relativistic dust, $w_{g} = 0$ and hence $c_{s}^2$ vanishes and the GCG clusters identically as CDM. But in the late times, when GCG starts behaving like DE, $w_{g} <0$. In this case $c_{s}^2$ is either positive or negative depending on the value of $\alpha$. And  due to this non-zero values for $c_{s}^2$ one gets unphysical oscillations or exponential blow-ups in the matter power spectrum. The only way to avoid this is to assume $\alpha \approx 0$ and in that case, GCG behaves exactly same as concordance $\Lambda$CDM model as mentioned above. This makes GCG indistinguishable from the concordance $\Lambda$CDM model.

Recently, Creminelli et al. \cite{crem} have shown that one can add higher derivative term like

\begin{equation}
{\cal L} _{1} = -\frac{M^{2}}{2} \left[ \Box \phi + 3 H(\phi)\right]^2,
\end{equation}

\noindent
where $M^{2} >0$, in the K-essence Lagrangian (1). It can be easily shown that addition of this term does not change the background evolution for the field or its energy density and pressure. But this term helps to keep the sound speed $c_{s}^2$ close to zero in order to have stability in the theory \cite{crem}.

As GCG equation of state also comes from a K-essence type field theory, we can add such a term in our Lagrangian (1) to keep the GCG equation of state unchanged in the background universe and at the same time we can keep the $c_{s}^2 \approx 0$ in the perturbed GCG fluid. 

This may help to remove the unphysical behaviour present in the matter power spectrum that would otherwise have occurred in the normal GCG fluid. In the next section we study this issue.

\section{Growth of Inhomogeneities}

Here we are interested in studying the growth of inhomogeneities in a clustering GCG fluid in the presence of baryons. We are mostly interested in small scale perturbations which are important to study the growth of structures in the universe. For this, Newtonian treatment is a valid approximation.  As we discuss in the previous section, for a clustering GCG  $c_{s}^2 =0$. We also have baryons which have similar vanishing sound speed. Therefore both the clustering GCG and baryons are comoving and hence they have the same peculiar velocity $v$. We follow the prescription by Sefusatti and Vernizzi \cite{vern} who have developed the system of equations for density perturbations for clustering quintessence in the presence of matter (baryons+CDM). We have the same system of equations where we replace clustering quintessence by clustering GCG and matter by baryons. Below we write down the perturbed continuity, Euler and Poisson equations \cite{vern}:

\begin{align}
&\frac{\partial \delta_b}{\partial \tau}  + \vec \nabla \cdot \big[ (1+\delta_b) \vec v \big] = 0 \label{continuity_m}\,,\\
&\frac{\partial \delta_g}{\partial \tau}  -3 w_{g} \mathcal{H} \delta_g+ \vec \nabla \cdot \big[ (1+w_{g}+\delta_g) \vec v\big] = 0 \label{continuity_Q}\,,\\
&\frac{\partial \vec v}{\partial \tau} + \mathcal{H}{\vec v} + ({\vec v}\cdot\nabla){\vec v} = -\nabla \Phi\,,\\
&\nabla^2 \Phi = 4 \pi G a^2  \left(\delta_g{\bar\rho}_g +  \delta_b{\bar\rho}_{b} \right).
\end{align}

\noindent
Here we write the equations in terms of conformal time $\tau$ and the density contrast $\delta_{\alpha} = \frac{\delta\rho_{\alpha}}{\bar{\rho_{\alpha}}}$ where $\delta\rho_{\alpha}$ and $\bar{\rho_{\alpha}}$ are the density perturbation and background energy density for the fluid ``$\alpha$". We further rewrite the Poisson's equation (3.4) as 

\begin{equation}
\nabla^2 \Phi = \frac{3}{2}{\mathcal{H}^2} \Omega_{b}(a) \left[\delta_{b} + \delta_{g}\frac{\Omega_{g}(a)}{\Omega_{b}(a)}\right]
\end{equation}

\noindent
and define the total density contrast as
\begin{equation}
\delta = \delta_{b} + \delta_{g}\frac{\Omega_{g}(a)}{\Omega_{b}(a)}.
\end{equation}

\begin{figure}
\begin{center}
\begin{tabular}{|c|c|}
\hline
 & \\
{\includegraphics[width=3in,height=3in,angle=0]{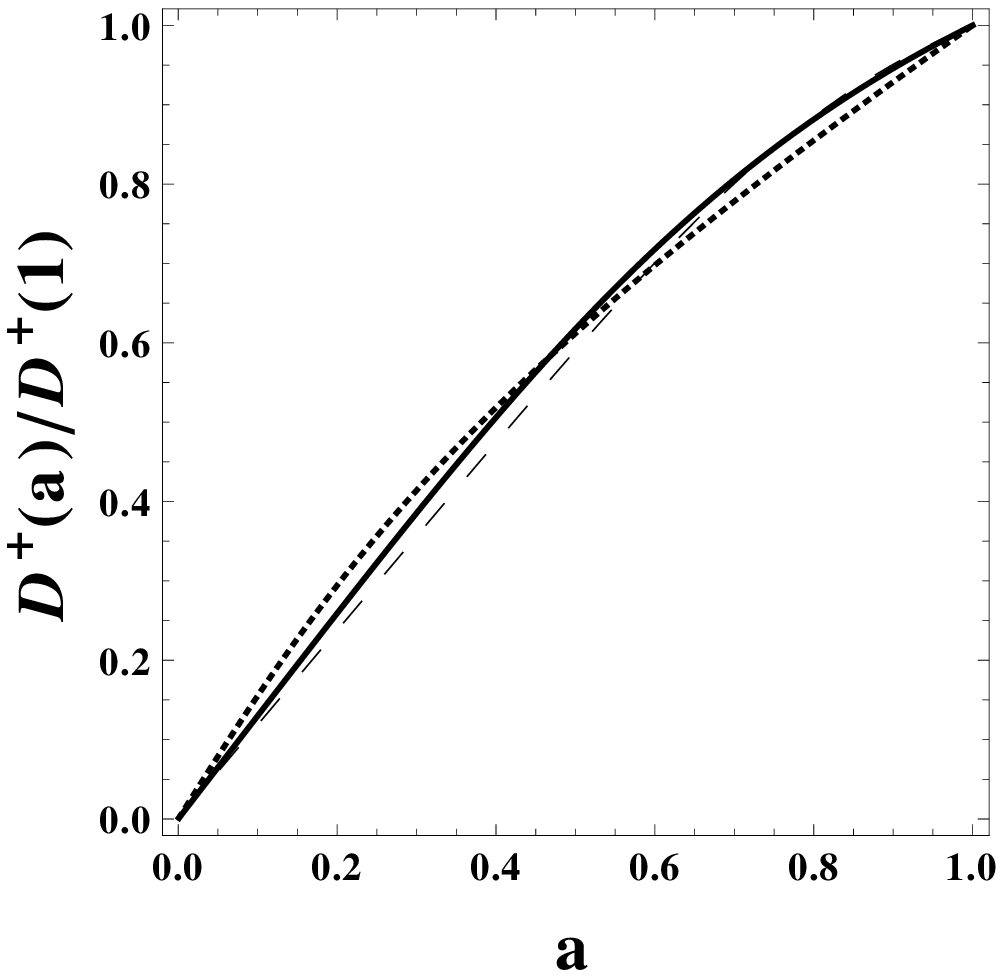}} &
{\includegraphics[width=3in,height=3in,angle=0]{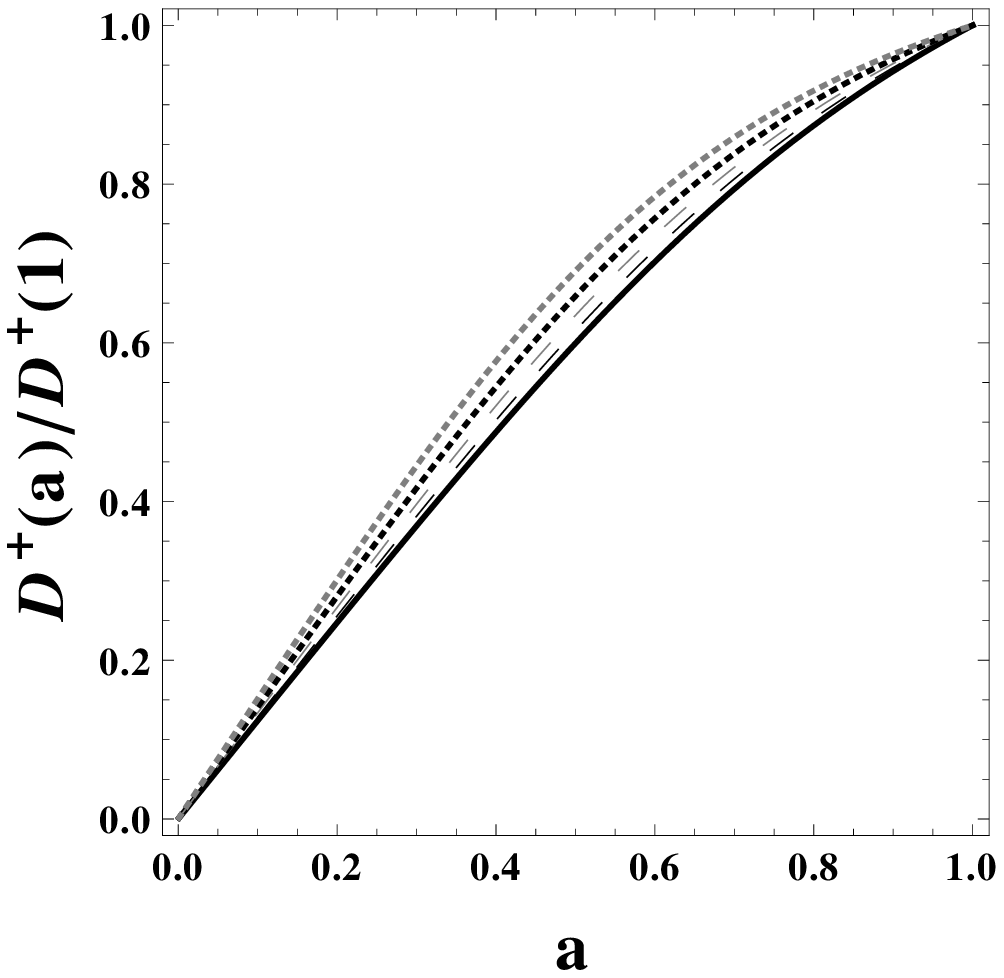}}
\\
\hline

\end{tabular}
\caption{Plot of $D^{+}(a)/D^{+}(1)$ vs $a$. Left Figure: $A_s = 0.75$. Solid, dashed and dotted curves represents $\alpha = 0.0, 0.9$ and $-0.9$ respectively. Right Figure:  $\alpha =0.1$.  $A_s = 0.7, 0.75, 0.8, 0.85$ and $0.9$ respectively from bottom to top. For both the figures we have taken $\Omega_{b0} = 0.045$}
\end{center}
\end{figure}

\noindent
Here $\Omega_{g} = (1-\Omega_{b})$, is the density parameter for clustering GCG. Next we study the solutions in the linear regime for the total density contrast $\delta$ defined in equation (3.5) by using equations (3.1), (3.2) and (3.5) by ignoring the higher order terms. Defining the linear growth function $D$  in the Fourier space and the linear growth rate $f$ as \cite{vern}

\begin{eqnarray}
\delta_{k}^{lin} (\tau) \equiv D(\tau)\delta_{k}^{lin};\\
f = \frac{d \ln{D}}{d \ln{a}},
\end{eqnarray}

\noindent
one can construct the equation for linear evolution for $D$ as

\begin{equation}
\frac{d^2 D}{d\ln a^2} + \left( \frac{1}{2} ( 1 - 3\, w_{g}\, \Omega_g)  -  \frac{d \ln C}{d \ln a} \right) \frac{d D}{d\ln a} - \frac32 \Omega_b  C D =0, \label{growth_function_evol}
\end{equation}

\noindent
where $C = 1 + (1 + w_{g}(a))\frac{\Omega_{g}(a)}{\Omega_{b}(a)}$. We should remind that the $\alpha = 0$ limit always reproduce the smooth $\Lambda$CDM result.  We solve the equation (3.9) from the period of decoupling ($a_{i} \sim 10^{-3}$) till today ($a=1$) with the initial conditions for the growing mode, $D^{+}(a) = a$ and $\frac{d D^{+}}{d a} \sim 1$. Subsequently we also calculate the growth rate $f$. The results are shown in Figure (1) and Figure (2) for different values of $\alpha$ and $A_{s}$. In each figure, we plot the behaviour for $\alpha=0$ $\Lambda$CDM case to compare with clustering GCG case. The figures show that the deviation from $\Lambda$CDM model is more prominent in the growth rate ($f$) evolution than in the growth function ($D^{+}$) evolution. Therefore it may not be possible to distinguish clustered GCG and $\Lambda$CDM using normal matter power spectrum which depends only on $D^{+}$, but measurements for redshift space distortion which depends on the growth rate $f$ can be useful to distinguish these two cases.

\begin{figure}
\begin{center}
\begin{tabular}{|c|c|}
\hline
 & \\
{\includegraphics[width=3in,height=3in,angle=0]{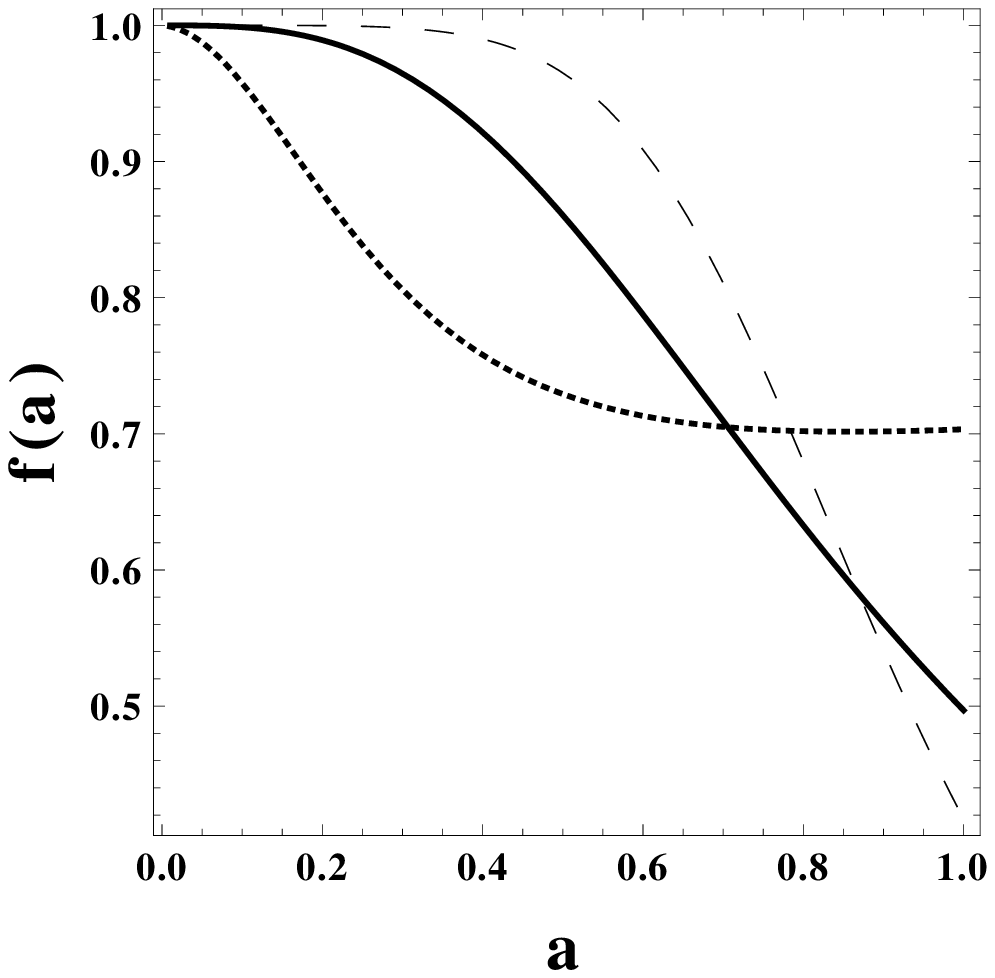}} &
{\includegraphics[width=3in,height=3in,angle=0]{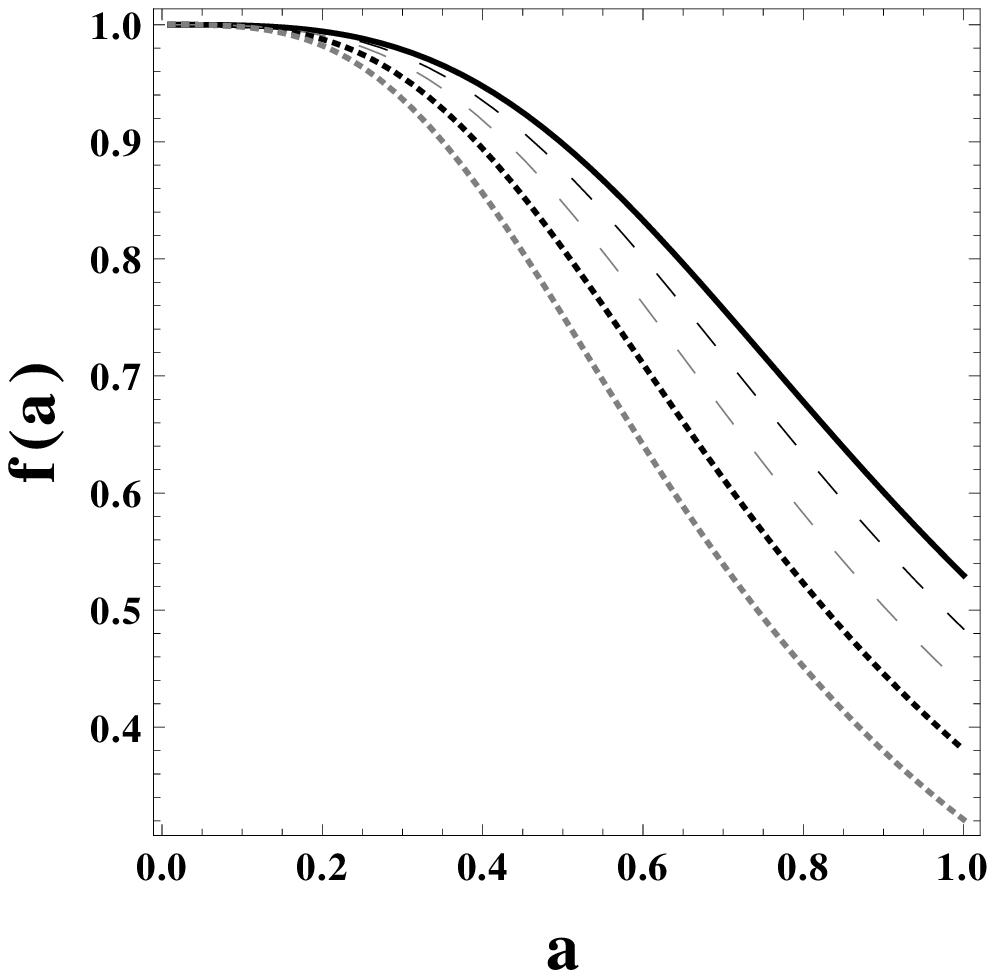}}
\\
\hline

\end{tabular}
\caption{Plot of $f(a)$ vs $a$. Left Figure:  $A_s = 0.75$. Solid, dashed and dotted curves represents $\alpha = 0.0, 0.9$ and $-0.9$ respectively. Right Figure: $\alpha= 0.1$. $A_s = 0.7, 0.75, 0.8, 0.85$ and $0.9$ respectively from top to bottom. For both the figures we have taken $\Omega_{b0} = 0.045$}
\end{center}
\end{figure}

\section{The power Spectrum}

The dimensionless linear power spectrum is defined as

\begin{equation}
\Delta^{2} (k,a) = \frac{k^{3} P(k,a)}{2\pi^2} = \delta_{H}^2 (\frac{k}{H_{0}})^{3+n_{s}}\left(\frac{D^{+}(a)}{D^{+}(1)}\right)^{2} T^{2} (k).
\end{equation}

\noindent
Here $\delta_{H}$ is the normalization constant, $H_{0}$ is the Hubble constant at present ($a=1$) and $n_{s}$ is spectral index for the primordial density fluctuations generated through inflation. In our case, $\Delta(k,a)$ represents the power spectrum for the total density contrast defined in equation (3.6). Now in the background universe GCG fluid behaves similar to non-relativistic matter except at the very late time when it starts behaving like a dark energy. On the other hand due to $c_{s}^{2} = 0$, the GCG clusters identically as CDM at all time and at all length scales. Hence it is safe to assume the transfer function $T(k)$ as prescribed by Eisenstein and Hu \cite{eisenhu}
for a mixture of CDM and baryons:

\begin{equation}
T(k) = \left(\frac{\Omega_{b0}}{\Omega_{m0}}\right)T_{b}(k) + \left(\frac{\Omega_{c0}}{\Omega_{m0}}\right)T_{c}(k)
\end{equation}

\noindent
where $\Omega_{c0}$ is density parameter for CDM and $\Omega_{m0} = \Omega_{c0}+\Omega_{b0}$. In our model, we identify $\Omega_{m0}$ with the model parameters as 

$$
\Omega_{m0} = \Omega_{b0}+\left(1-A_{s}\right)^{\frac{1}{1+\alpha}}\left(1-\Omega_{b0}\right)
$$

The form for $T_{b}(k)$ and $T_{c}(k)$ are given by Eisenstein and Hu \cite{eisenhu}. One important quantity related to the growth of structures is the variance $\sigma_{8}$ of the density fluctuation at the scale $8 h^{-1}$ Mpc. This is defined as

\begin{equation}
\sigma^{2}(a,R) = \int^{\infty}_{0} \Delta^{2}(k,a) W^{2}(k,R) d \ln{k},
\end{equation}

\noindent where the window function $W(k,R)$  is defined as $W(k,R) = 3\left( \frac{\sin(KR)}{(kR)^3} + \frac{\cos(kR)}{(kr)^2}\right)$ and for $\sigma_{8}$, one puts $ R = 8h^{-1}$ Mpc.

\section{Observational Constraints}

In this section, we confront the clustering GCG model with various observational data. Previous results for GCG as UDME showed that in order to satisfy various observational constraints, specially those coming from the inhomogeneous universe, GCG has to be indistinguishable from the $\Lambda$CDM model. Here we want to see, with clustering GCG as UDME, whether we have a allowed parameter space for GCG where it behaves differently from $\Lambda$CDM model. Below we describe the various observational data that we use to put constraint on the clustering GCG model.

\begin{figure}
\centering
{\includegraphics[width=2.5in,height=2in,angle=0]{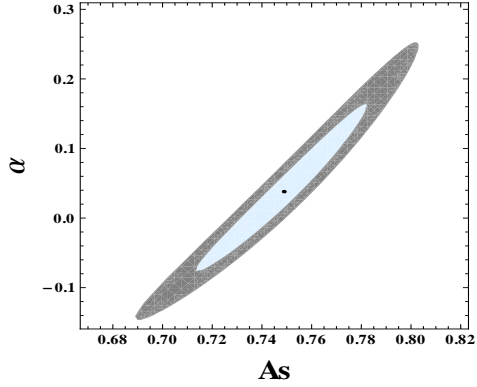}} 
\caption{The $1\sigma$ and $2\sigma$ confidence regions in $A_s-\alpha$ plane. We have used SN+Hubble+BAO+ $f\sigma_{8}$ data. The black dot represents the best fit values for $A_{s}$ and $\alpha$ (see text).}
\end{figure}


\begin{itemize}

\item We consider the Union2.1 compilation containing 580 data points for distance modulus $\mu$ for Type-Ia supernova at various redshift \cite{sn}.

\item We use the compilation of 28 observational data points at various redshifts for the Hubble parameters by Farooq and Ratra using different evolutions of cosmic chronometers within the redshift range $0.07 < z < 2.3$ \cite{ratra}. This spans almost 10 Gyr of cosmic evolutions.

\item We use the combined BAO/CMB constraints on the angular scales of the BAO oscillations in the matter power spectrum as measured by SDSS survey, 6dF Galaxy survey and the Wiggle-z survey. The full covariance matrix for this has been provided by Giostri et al \cite{bao}.

\item Finally we use the measurements for $f\sigma_{8}$ by various Galaxy surveys like 2dF, SDSS, 6dF, BOSS and Wiggle-Z. This combination is an excellent model  independent estimator of the growth of structure formations and has been measured by various surveys. In a recent paper by Basilakos et al. a compilation of different measurements for $f\sigma_{8}$ has been provided \cite{sigma8} and we use that compilation for our purpose.
\end{itemize}

\begin{figure}[t]
\centering 
\includegraphics[scale=0.88]{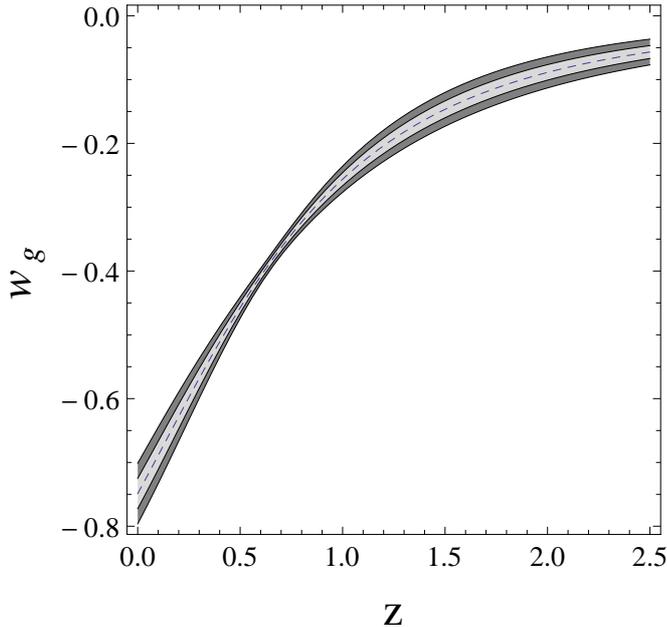}
\caption{ Allowed behaviour at $1\sigma$ and $2\sigma$ confidence level for the equation of state for the GCG, $w_{g}$, as a function of redshift.}
\end{figure}

\begin{figure}[t]
\centering 
\includegraphics[scale=0.88]{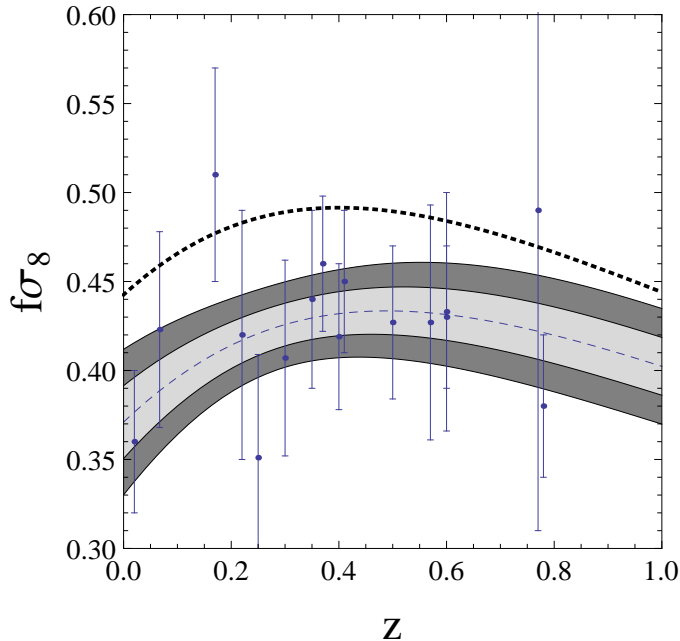}
\caption{ The $1\sigma$ and $2\sigma$ allowed band for $f\sigma_{8}$ evolution with redshift. The dashed line is for the best fit values for $A_{s}$ and $\alpha$ for GCG UDME model. The dotted line is for the $\Lambda$CDM best fit model as obtained by Planck experiment \cite{planck}. This specific line for $\Lambda$CDM as well as the band and the best fit line for GCG UDME are scale independent (see text for explanation)The data points are the measurements for $f\sigma_{8}$ from various experiments as compiled in \cite{sigma8}. These are the data that we also use in our analysis.}
\end{figure}

Using the above observational data, we put constraint in the $\alpha-A_{s}$ parameter space fixing all the cosmological parameters at best fit values obtained by Planck experiment \cite{planck}. These are as follows: $\Omega_{b0} = 0.045$, $n_{s} = 0.96$ and $H_{0}=67.04 km s^{-1} Mpc^{-1}$.  We also normalize the matter power spectrum (4.1) using $\sigma_{8} = 0.8347$ as obtained by Planck \cite{planck}. 
Using all the data that we describe above one gets the  best fit values for $A_{s}$ and $\alpha$ as $ 0.75$ and $0.043$ respectively and the $1\sigma$ error bars as $0.75\pm 0.023$ and $ 0.043\pm 0.079$ respectively.

The constrained region in the $A_{s}-\alpha$ parameter space for the combined data is shown in Figure (3). It shows that although $\alpha=0$ which represents the $\Lambda$CDM model, is allowed, the combined data also allows non zero values for $\alpha$ which represents UDME. Also the constraint on $\alpha$ is reasonably tight and the data rule out the $\alpha=1$ original CG behaviour. We should also point that wiggleZ data alone does not contribute much in constraining the $A_{s}-\alpha$ parameter space.

Next, using the covariance matrix for $\alpha$ and $A_{s}$ that we obtain during the data analysis, and together with the standard error propagation technique, we calculate the error in the equation of state $w_{g}$ for the clustering GCG. The result is shown in the Figure (4). The figure shows that the variation of the equation of state for the clustering GCG with redshift is very tightly constrained. The constraint is even tighter around redshift $z=0.6$. 

In figure (5), we show the $1\sigma$ and $2\sigma$ allowed region for evolution for the combination $f\sigma_{8}$ as a function of redshift $z$. We stress that the shaded region in this figure represents the observational constraint for $f\sigma_{8}$ as a function of redshift for the GCG UDME model. This has been obtained from the covariance between $A_{s} - \alpha$ using combined datasets mentioned earlier in the text. We also show the same evolution for the  best fit $\Lambda$CDM model as obtained by Planck\cite{planck}. Note that this specific line for $\Lambda$CDM model is scale independent as pressureless matter clusters at all scales equally due to vanshing sound speed. This is also true for our clustering GCG UDME which also has $c_{s}^2 = 0$. It is evident that our result is not fully consistent with the best fit $\Lambda$CDM model obtained by Planck. Remember that we only use the low redshift measurements for all the observables while Planck is related to CMB measurement at redshift $z \sim 1000$. This discrepancy in the $f\sigma_{8}$ measurement from low and high redshift data was also reported earlier by Macaulay et al. \cite{erik}.

And finally we show the behaviour of the matter power spectrum  in Figure (6) as defined in equation (4.1) using  combinations for $A_{s}$ and $\alpha$ which are allowed at $1\sigma$ and $2\sigma$ confidence levels. We also show the $\Lambda$CDM  ($\alpha = 0$) case. As can be seen from this plot, there is no unphysical oscillations or exponential blow-ups for models different from $\Lambda$CDM. The difference between the clustering GCG and $\Lambda$CDM is also shown. For a model which is allowed at $2\sigma$ confidence limit, one can see a $20\%$ deviation from the $\Lambda$CDM which is quite substantial.

\section{Conclusion}
In this paper, we study the possibility of having a unified model  for dark matter and dark energy (UDME) using the clustering GCG where by incorporating higher dimensional operator in the original K-essence Lagrangian for GCG, we keep the $c_{s}^2 \approx 0$ in the GCG fluid. This original idea was put forward by Creminelli et al. \cite{crem} and we apply it to GCG. By doing so, we ensure GCG clusters at all scales similar to the CDM. As in the background universe, GCG behaves like a CDM in the early time and like DE in the late time, this added clustering property makes GCG a suitable candidate for the UDME.

We study the growth of density fluctuations in the linear regime in the clustering GCG model. We show that the growth of density fluctuations in the clustering GCG can deviate appreciably from the $\Lambda$CDM behaviour. This is most prominent in the behaviour of the linear growth rate $f$.  

Subsequently using the recent observational data from SnIa, H(z), BAO measurements and the measurements of $f\sigma_{8}$ by various galaxy surveys, we put constraint on the model parameters $\alpha$ and $A_{s}$. We get a tight constraint on the model parameter $\alpha$. The $\alpha=0$,  $\Lambda$CDM model is allowed but the original $\alpha=1$ CG model is ruled out. The constraint on the variation of the GCG equation of state $w_{g}$ with redshift is extremely tight. We also obtain the allowed variation for the $f\sigma_{8}$ combination with redshift at $1\sigma$ and $2\sigma$ confidence limit. The results shows a discrepancy with the best fit $\Lambda$CDM model as obtained by Planck. We also show behaviour of the matter power spectrum confirming the absence of any unphysical behaviour. The deviation from the $\Lambda$CDM model can be as high as $20\%$ for clustering GCG models which are consistent with the data at $2\sigma$ confidence level.

\begin{figure}
\begin{center}
\begin{tabular}{|c|c|}
\hline
 & \\
{\includegraphics[width=2.5in,height=2in,angle=0]{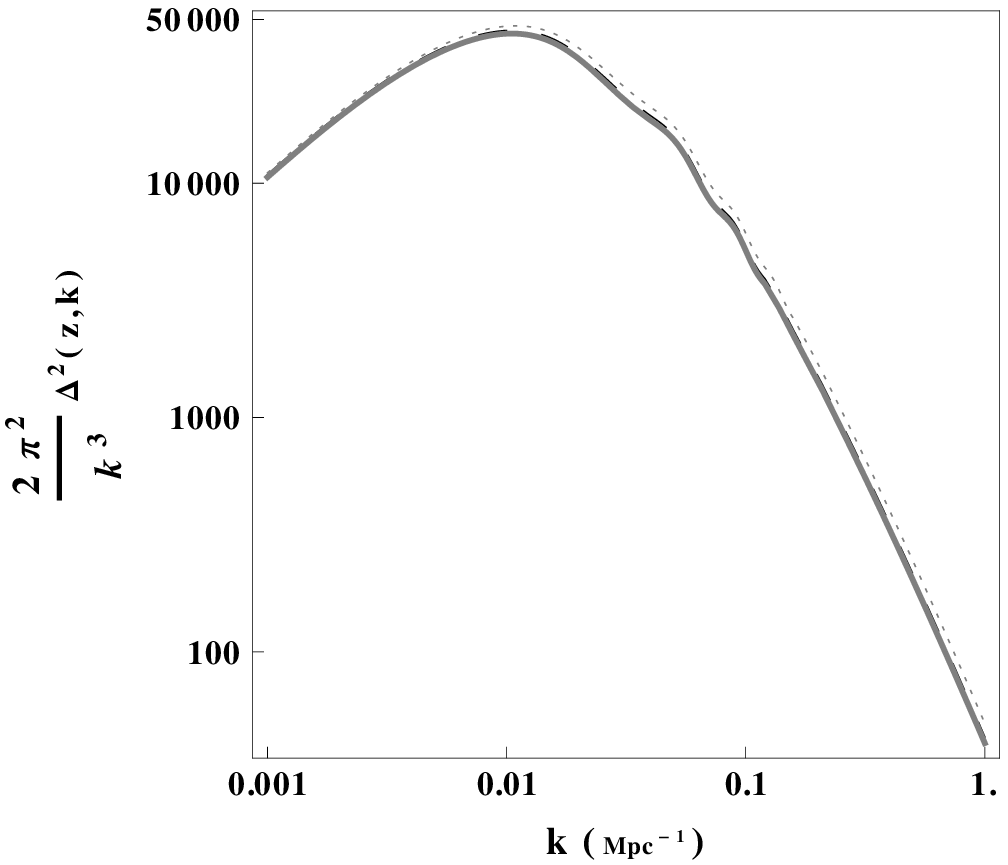}} &
{\includegraphics[width=2.5in,height=2in,angle=0]{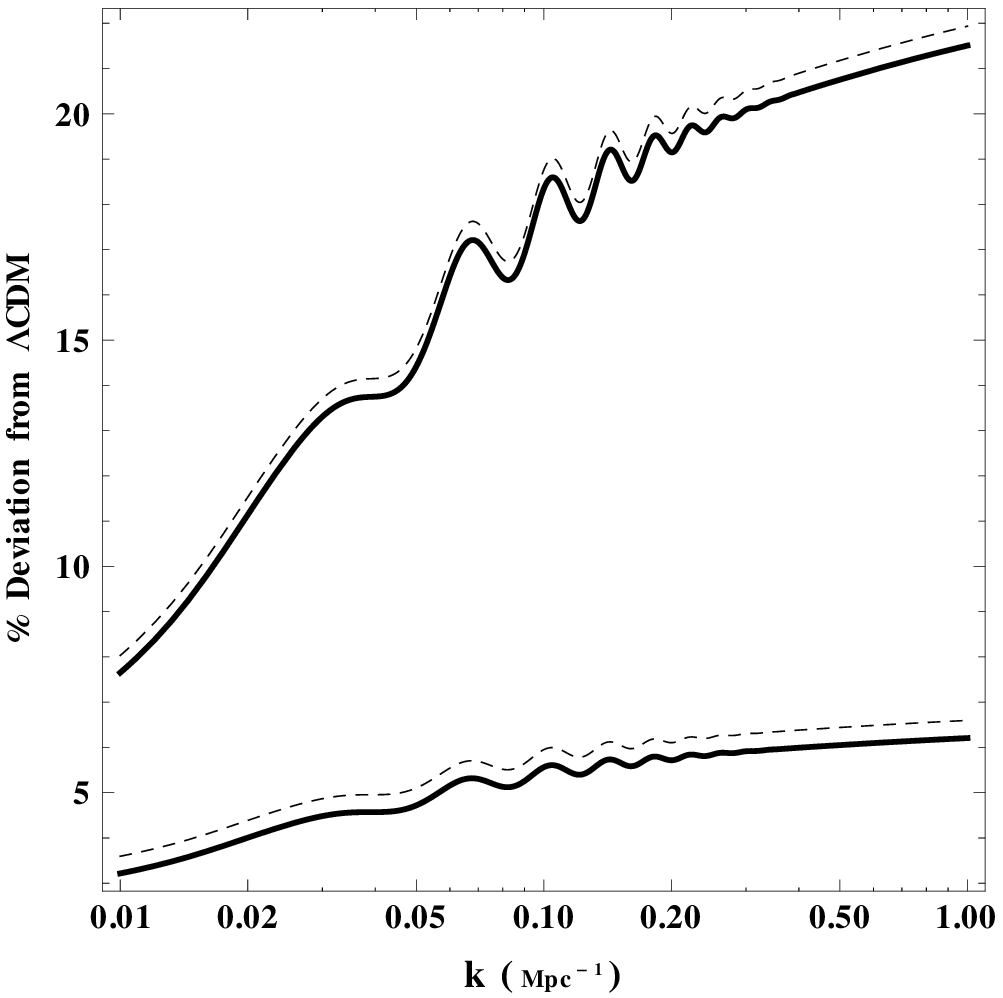}}
\\
\hline

\end{tabular}
\caption{{\it Left Figure} : The behaviour of the power spectrum  at $z=0.5$ for different $A_{s}$ and $\alpha$ combinations that are allowed by the joint dataset (see text). Solid line is for $\alpha = 0$ ($\Lambda$CDM) and $A_{s} = 0.74$. The dashed line is for $A_{s} =0.77$ and $\alpha = 0.1$ (allowed at $1\sigma$).  The dotted line is for $A_{s} =0.79$ and $\alpha = 0.2$ (allowed at $2\sigma$). {\it Right Figure}: The percentage deviation for the power spectrum from $\Lambda$CDM model. The upper one is for  $A_{s} =0.79$ and $\alpha = 0.2$. The lower one is for $A_{s} =0.77$ and $\alpha = 0.1$. For each of these sets, the solid is for $z=0.5$ and the dashed line is for $z=1.0$. }
\end{center}
\end{figure}

After the initial enthusiasm for GCG as UDME, the interest gradually decayed due to the unwanted features it produces in the matter power spectrum although as a DE parameterization, GCG is still an interesting option (see \cite{hazra} for the recent study). After this study, we believe there will be renewed interests in GCG as a viable option for UDME.  

\section*{Acknowledgments}
A.A.S. acknowledges the funding from SERC, Dept. of Science and Technology, Govt. of India through the research project SR/S2/HEP-43/2009.  S.K. thanks the UGC, Govt. of India for financial support.

\end{document}